\documentclass[12pt]{article}

\usepackage{graphicx}
\usepackage{amsfonts}
\usepackage{amssymb}
\usepackage{amsbsy}

\input epsf.sty

\newcommand{\BEQ}{\begin{equation}}     
\newcommand{\BEA}{\begin{eqnarray}}
\newcommand{\EEQ}{\end{equation}}       
\newcommand{\EEA}{\end{eqnarray}}
\newcommand{\eps}{\varepsilon}          
\newcommand{\sig}{\sigma}               
\newcommand{\D}{{\rm d}}                
\newcommand{\II}{{\rm i}}               
\newcommand{\wit}[1]{\widetilde{#1}}    
 
\renewcommand{\vec}[1]{\boldsymbol{#1}} 

\newcommand{\vekz}[2]
     {\mbox{${\begin{array}{c} #1  \\ #2 \end{array}}$}}

\begin{document}

\begin{titlepage}


\vskip 1.5 cm
\begin{center}
{\Large \bf Local scale invariance and its application to strongly
anisotropic critical phenomena}
\end{center}

\vskip 2.0 cm
   
\centerline{{\bf Malte Henkel}$^a$, {\bf Alan Picone}$^a$, 
{\bf Michel Pleimling}$^b$ and {\bf J\'er\'emie Unterberger}$^c$ }
\vskip 0.5 cm
\centerline{$^a$Laboratoire de Physique des 
Mat\'eriaux\footnote{Laboratoire associ\'e au CNRS UMR 7556} and}
\centerline{
Laboratoire Europ\'een de Recherche Universitaire Sarre-Lorraine,}
\centerline{ 
Universit\'e Henri Poincar\'e Nancy I, B.P. 239,} 
\centerline{ 
F -- 54506 Vand{\oe}uvre l\`es Nancy Cedex, France\\}

\centerline{$^b$Institut f\"ur Theoretische Physik I, 
Universit\"at Erlangen-N\"urnberg, D -- 91058 Erlangen, Germany\\}

\centerline{$^c$Institut \'Elie Cartan,\footnote{Laboratoire associ\'e au 
CNRS UMR 7502} D\'epartement de Math\'ematiques, 
Universit\'e Henri Poincar\'e Nancy I, B.P. 239,} 
\centerline{ 
F -- 54506 Vand{\oe}uvre l\`es Nancy Cedex, France}

\begin{abstract}
\noindent 
The generalization of dynamical scaling to local scale invariance is
reviewed. Starting from a recapitulation of the phenomenology of ageing
phenomena, the generalization of dynamical scaling to local scale 
transformation for any given dynamical exponent $z$ is described and the
two distinct types of local scale invariance are presented. The 
special case $z=2$ and the associated Ward identity of Schr\"odinger 
invariance is treated. Local scale invariance predicts the form of the
two-point functions. Existing confirmations of these predictions for
(I) the Lifshitz points in spin systems with competing interactions such as
the ANNNI model and (II) non-equilibrium ageing phenomena as occur in the
kinetic Ising model with Glauber dynamics are described.  
\end{abstract}
\end{titlepage}

\section{Ageing and dynamical scaling}

Non-equilibrium critical phenomena are a subject of intense research 
activity in physics. Rather than providing a general and exhaustive
definition, we shall present a typical example of current interest, chosen
such that our main question can be introduced in a natural way.
 
A common way to reach a situation of non-equilibrium criticality 
is through a rapid change of one of the macroscopic variables which enter 
into the equation of state. For definiteness, consider a simple ferromagnet. 
The simplest model system of this kind is the {\em Ising model} which considers
magnetic moments attached to the sites of a (regular) lattice $\Lambda$. 
By hypothesis, these moments may only take two values, say `up' and `down', 
and are described
through a spin variable $\sig_{\vec{i}}=\pm 1$ attached to the lattice site
$\vec{i}\in\Lambda$. In equilibrium, to a given configuration of spins 
$\{\sig\}=\{\sig_{\vec{i}}\}_{\vec{i}\in\Lambda}$
one associates an energy 
${\cal H}\{\sig\} = -\sum_{(\vec{i},\vec{j})} \sig_{\vec{i}}\sig_{\vec{j}}$ 
where the sum is usually restricted to nearest-neighbour pairs
on the lattice. The probability of a given configuration $\{\sig\}$ in 
equilibrium is then given by 
\BEQ
P_{\rm eq}(\{\sig\}) = \frac{1}{Z} \exp \left(- {\cal H}\{\sig\}/T\right) 
\;\; , \;\;
Z := \sum_{\{\sig\}} \exp \left(- {\cal H}\{\sig\}/T\right) 
\EEQ
where $T$ is the temperature. 
It is well-known that systems of this kind show an order-disorder 
phase transition, at least in $d>1$ dimensions, such that the {\em order
parameter} $\phi :=\langle\sum_{\vec{i}\in\Lambda}\sig_{\vec{i}}\rangle$ 
has a non-vanishing value for temperatures below a 
certain critical temperature $T_c>0$ and vanishes
for $T>T_c$. Such a model may be brought out of equilibrium, for example, 
by starting initially from a fully disordered state (effectively corresponding 
to an infinite temperature) and then quench the system 
rapidly to a fixed temperature below the system's critical temperature $T_c>0$. 
Empirically, it is then found that the system evolves slowly towards the
equilibrium state corresponding to the fixed temperature $T$. However, the
details of this relaxation process may be complex. In many materials 
undergoing such thermal treatment, the time-dependent properties may well 
depend on the entire thermal history of the sample. 
The resulting {\em ageing} behaviour has been in the focus of intensive study, 
see \cite{Stru78,Bray94,Cate00,Godr02,Cugl02} for reviews.

In the Ising model example, a popular choice to simulate the update
of the individual spins is through Glauber dynamics \cite{Glau63}. 
Glauber dynamics may be 
realized through the discrete-time 
heat-bath rule $\sig_{\vec{i}}(t)\to\sig_{\vec{i}}(t+1)$ such that
\BEQ \label{7:gl:heat}
\sig_{\vec{i}}(t+1) = \pm 1 \mbox{\rm ~~ with probability $\frac{1}{2}
\left[ 1\pm \tanh(h_{\vec{i}}(t)/T)\right]$}
\EEQ
with the local field 
$h_{\vec{i}}(t)=\sum_{\vec{j}(\vec{i})}\sig_{\vec{j}}(t)$, where
$\vec{j}(\vec{i})$ runs over the nearest neighbours of the site $\vec{i}$. 
The main property
of this dynamical rule is that the order parameter $\phi$ is {\em not}
conserved. For non-equilibrium systems, there is no general theoretical 
framework for the calculation of the time-dependent probabilities
$P(\{\sig\};t)$. However, with the choice
(\ref{7:gl:heat}), $P(\{\sig\};t)$ can be found exactly in $1D$ by solving an
associated master equation \cite{Glau63}. The time-dependent spin-spin 
correlators and their approach towards equilibrium can be determined exactly
in this special case. Otherwise, numerical simulation must be used.  

\begin{figure}[t]
\centerline{\epsfxsize=2.3in\ \epsfbox{
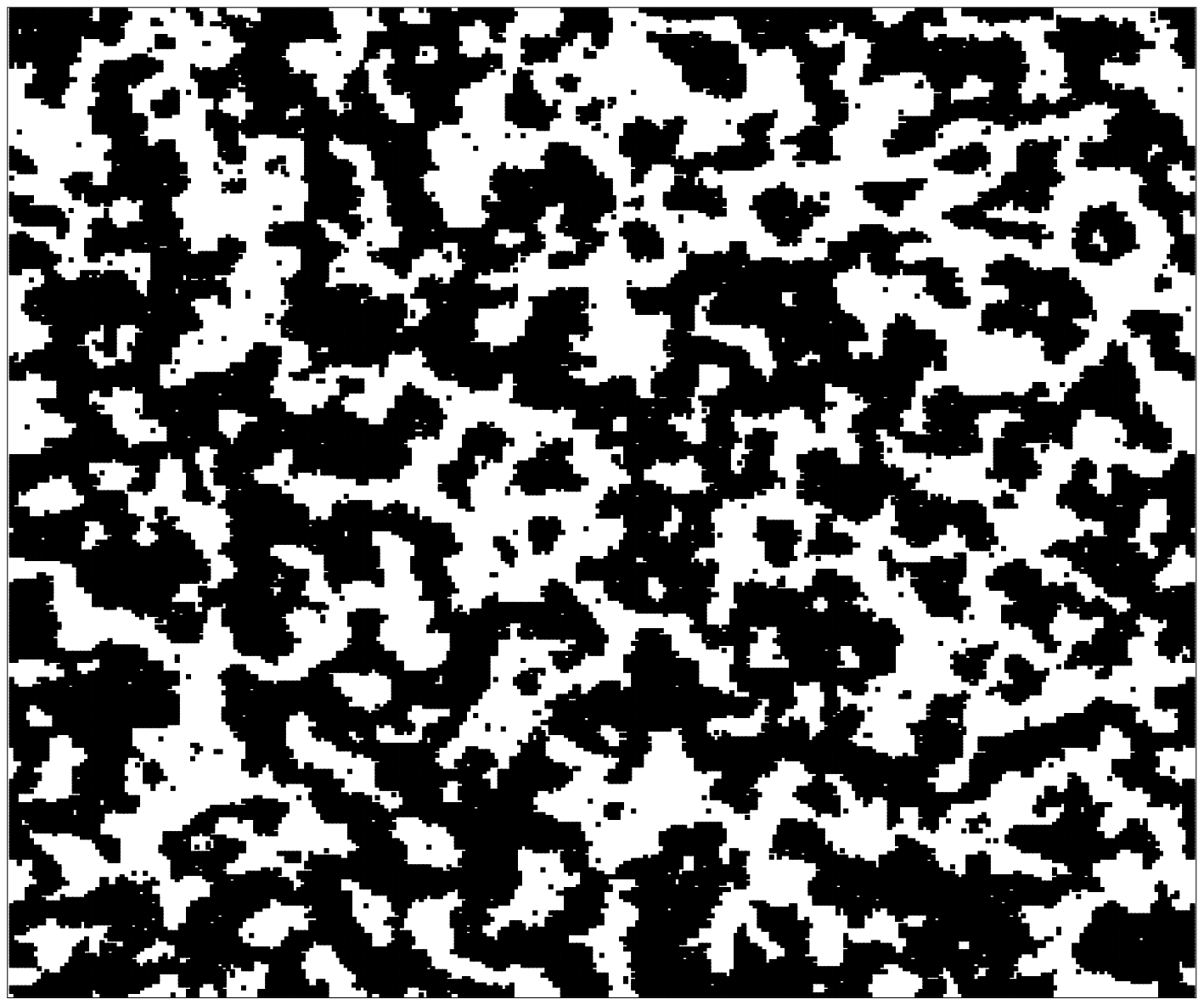} ~~
\epsfxsize=2.3in\epsfbox{
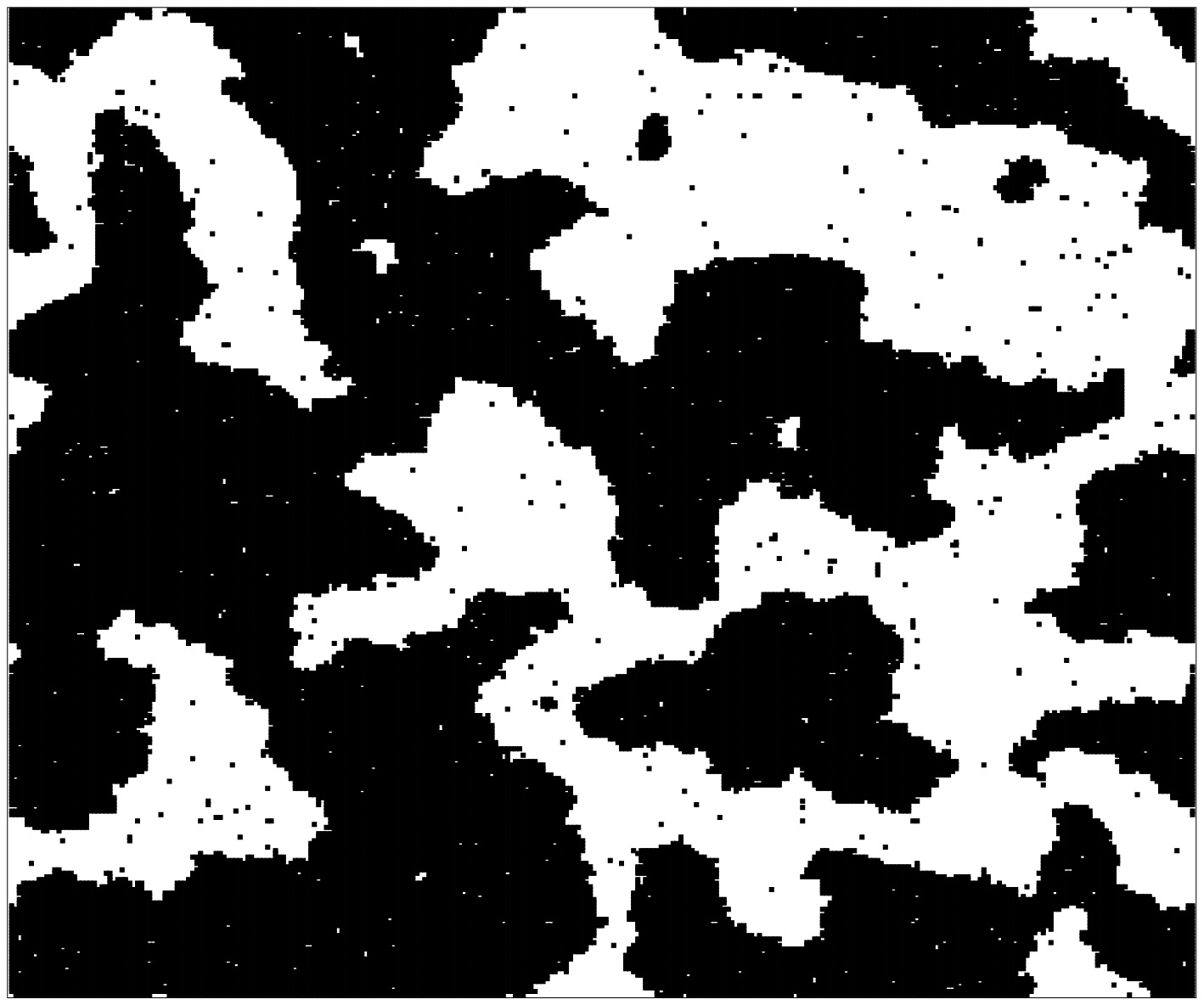}
}
\caption[Coarsening]{Snapshots of the coarsening of ordered domains in the 
$2D$ Glauber-Ising model, after a quench to $T=1.5<T_c$ from a totally 
disordered state and at times $t=25$ (left) and $t=275$ (right) after the 
quench.
\label{Bild_1}}
\end{figure}

In figure~\ref{Bild_1}, we illustrate the evolution of the microscopic
state after a quench to $T<T_c$. We see that ordered domains form and slowly
grow, with a typical size $L(t) \sim t^{1/z}$, where $z$ is the dynamical
exponent. The presence of {\em algebraic} growth laws for such dynamic 
quantities, although the equilibrium state of the model is not critical, 
is characteristic for the ageing phenomena we wish to consider. Indeed, inside
a given domain the spins are completely ordered and the time evolution of the
model only occurs through the slow motion of the domain walls.

This evolution is more fully revealed through the
study of {\em two-time} quantities, such as the two-time correlation
function $C(t,s;\vec{r})$ and the (linear) 
autoresponse function $R(t,s;\vec{r})$
\BEQ
C(t,s;\vec{r}) = \langle \phi(t,\vec{r}) \phi(s,\vec{0}) \rangle \;\; , \;\;
R(t,s;\vec{r}) = \left.
\frac{\delta\langle\phi(t,\vec{r})\rangle}{\delta h(s,\vec{0})}\right|_{h=0}
\EEQ
where $\phi$ is the order parameter, $h$ the conjugate magnetic field,
$t$ is called the observation time and $s$ the waiting time. Space translation
invariance is assumed throughout. Autocorrelators
are given by $C(t,s)=C(t,s;\vec{0})$ and autoresponses by
$R(t,s)=R(t,s;\vec{0})$.\footnote{We
implicitly assume here that the mean order parameter $\langle\phi(t)\rangle=0$,
otherwise $C(t,s)$ must be replaced by the connected correlator
$\Gamma(t,s)=C(t,s)-\langle\phi(t)\rangle\langle\phi(s)\rangle$.} One says
that a system is {\em ageing} when $C(t,s)$ and/or $R(t,s)$ depend on both
$t$ and $s$ and not only on their difference. Ageing phenomena are
therefore related to a {\em breaking of time-translation invariance}. 
Physically, ageing occurs in the regime when $s$ and $\tau=t-s>0$ are 
simultaneously much larger than any microscopic time scale $\tau_{\rm micro}$. 
In many systems, one finds in the ageing regime a dynamical scaling behaviour, 
see \cite{Cate00,Godr02} 
\BEQ \label{gl:CR}
C(t,s) = s^{-b} f_C(t/s) \;\; , \;\; R(t,s) = s^{-1-a} f_R(t/s)
\EEQ
where $a$ and $b$ are non-equilibrium exponents. 
We emphasize that this non-equilibrium scaling also occurs for
quenches exactly to the critical point $T=T_c$, but the values of the
exponents found at $T=T_c$ will be different from those found for $T<T_c$. 
For $T<T_c$, it can be shown from phenomenological arguments that $z=2$
for dynamical rules with a non-conserved 
order parameter \cite{Rute95,Bray94}, while
at $T=T_c$, $z$ has a non-trivial value which must be determined from a 
full renormalization-group approach. 

\begin{table}
\caption{Non-equilibrium exponents $a$ and $b$ for quenches from a fully
disordered state onto and below
the critical point $T_c>0$ of simple ferromagnets, 
according to \cite{Henk02a}. \label{Tabelle1}}
\begin{center}
\begin{tabular}{|c|cc|c|}  \hline 
        & $a$            & $b$            & Class \\ \hline
$T=T_c$ & $(d-2+\eta)/z$ & $(d-2+\eta)/z$ &   \\ \hline
$T<T_c$ & $(d-2+\eta)/z$ & 0              & L \\ 
        & $1/z$          & 0              & S \\ \hline 
\end{tabular}
\end{center}
\end{table}

In table~\ref{Tabelle1} we collect the values of the non-equilibrium exponents
$a$ and $b$. Indeed, it turns out that for $T<T_c$ the equilibrium behaviour 
of the order parameter correlator is essential. If 
$C_{\rm eq}(\vec{r})\sim e^{-|\vec{r}|/\xi}$ with a finite $\xi$, we say that
the system is of {\em class S}, while if 
$C_{\rm eq}(\vec{r})\sim |\vec{r}|^{-(d-2+\eta)}$, the system is said to 
be of {\em class L}, where $\eta$ is a standard equilibrium critical exponent.
The results for $a$ are based on the well-accepted 
physical image that ageing comes from
the slow motion of the domain walls \cite{Bray94,Cate00,Bert99}. This idea
had been questioned recently and it was argued that because of some anomalous
scaling behaviour, the simple relation $a=1/2$ should be invalid in the
$2D$ Glauber-Ising model (which belongs to class S) and rather $a=1/4$
here \cite{Corb02}. However, 
recent tests have refuted this claim and reestablish that indeed $a=1/2$
in this model \cite{Henk02a,Henk03b}. 

The scaling functions are expected to 
behave for large arguments $x=t/s\gg 1$ asymptotically as
\BEQ
f_C(x) \sim x^{-\lambda_C/z} \;\; , \;\; f_R(x) \sim x^{-\lambda_R/z}
\EEQ
where $\lambda_C$ and $\lambda_R$ are called 
the {\em autocorrelation exponent} \cite{Fish88,Huse89} 
and {\em autoresponse exponent} \cite{Pico02}, respectively. For a fully
disordered initial state, it is traditionally accepted that 
$\lambda_C=\lambda_R=\lambda$. However, for spatially long-ranged 
initial correlations of
the form $C_{\rm ini}(\vec{r})\sim |\vec{r}|^{-d-\alpha}$ (with $\alpha\leq 0$)
the relation 
\BEQ \label{gl:lVerm}
\lambda_C=\lambda_R+\alpha
\EEQ 
has been conjectured \cite{Pico02} and $\lambda_C=\lambda_R$ is only recovered
if $\alpha$ renormalizes to zero.\footnote{The conjecture (\ref{gl:lVerm}) 
is in agreement with all known results from the kinetic spherical model
\cite{Pico02,Newm90} and also explains those of the $2D$ XY model with 
$T\leq T_{\rm KT}$ \cite{Bert01}. Remarkably, for the Ising model with
long-range initial conditions, only $\lambda_C$ appears to be known
\cite{Huma91}.} 
Furthermore, the rigorous arguments of \cite{Yeun96} readily yield
$\lambda_C\geq (d+\alpha)/2$, for simple ferromagnets without disorder. 
Very recently, distinct exponents 
$\lambda_C\ne\lambda_R$ have also been found in the {\em random} 
sine-Gordon model and in addition $\lambda_C< d/2$ violates the rigorous bound 
mentioned above \cite{Sche03}. 

Another central question in this context is 
whether/when under the conditions just described the system is in
thermodynamic equilibrium. It is convenient to consider 
the fluctuation-dissipation ratio \cite{Cugl94a,Cugl94b}
\BEQ
X(t,s) = T R(t,s) \left( \frac{\partial C(t,s)}{\partial s}\right)^{-1}
\label{1:eq5}
\EEQ
At equilibrium, the fluctuation-dissipation theorem states that $X(t,s)=1$.
The breaking of the fluctuation-dissipation theorem has been investigated 
intensively both theoretically 
(see e.g. \cite{Cate00,Godr02,Garr01,Pere02,Cugl02,Cris03}) 
and experimentally \cite{Grig99,Heri02,Bell02}.

Having thus reviewed the main phenomenological aspects of the ageing
behaviour of simple ferromagnets, we can now formulate our main question: 
{\it is it possible to generalize the dynamical scaling described by
eq.~(\ref{gl:CR}) to a local scale invariance~?} By this we mean that
we wish to generalize the transformations of dynamical scaling 
$t\mapsto  b^z t$ and $\vec{r}\mapsto b \vec{r}$ to a local form where
$b=b(t,\vec{r})$ \cite{Henk02}. We therefore seek to extend dynamical scaling 
in a way analogous to the extension of global scale invariance in 
equilibrium critical phenomena to conformal invariance. 

This review is organized as follows. In section 2 we present the
main points of the construction of local scale transformations, for
an arbitrary dynamical exponent $z$. In section 3 we consider in more detail
the case $z=2$ where the relation to the conformal group and local Ward
identities are discussed. In section 4, general expressions for the scaling
functions of two-point functions are derived. Application to Lifshitz
points are briefly discussed in section 5. 
Finally, in section 6 we return to applications
of local scale invariance to ageing systems, focussing in particular on the 
prediction of the scaling of the response functions and tests in the 
Glauber-Ising model. 

\section{Construction of local scale transformations}

We are interested in systems with strongly anisotropic
or dynamical criticality. By definition, two-point functions of such systems 
satisfy the scaling form
\begin{equation} \label{gl:skala}
G(t,\vec{r}) = b^{2x} G(b^{\theta}t, b\vec{r}) = t^{-2x/\theta} \Phi\left(
r t^{-1/\theta}\right) = r^{-2x} \Omega\left( t r^{-\theta}\right)
\end{equation}
where $t$ stands for `temporal' and $\vec{r}$ for `spatial' coordinates,
$x$ is a scaling dimension, $\theta$ the anisotropy exponent (when $t$ 
corresponds to physical time, $\theta=z$ is called the dynamical exponent) and
$\Phi,\Omega$ are scaling functions. Physical realizations of this are numerous,
see \cite{Stru78,Bray94,Cate00,Cugl02,Card96} and references therein. 
For isotropic critical systems,
$\theta=1$ and the `temporal' variable $t$ becomes just another coordinate. 
It is well-known that in this case, scale invariance (\ref{gl:skala}) with
a constant rescaling factor $b$ can be replaced by the larger group 
of conformal
transformations $b=b(t,\vec{r})$ such that angles are preserved. It turns
out that in the case of one space and one time dimensions, conformal invariance
becomes an important dynamical symmetry from which many physically relevant 
conclusions can be drawn \cite{Bela84}. 

Given the remarkable success of conformal invariance descriptions of 
equilibrium phase transitions, see e.g. \cite{Henk99,diFr97} for introductions, 
one may wonder whether similar extensions of
scale invariance also exist when $\theta\neq 1$. Indeed, for $\theta=2$ the
analogue of the conformal group is known to be the Schr\"odinger group
\cite{Nied72,Hage72} (and apparently already known to Lie). 
We shall first describe the construction of these
{\em local scale transformations} for arbitrary $\theta\ne 1,2$, 
show that they act as a dynamical
symmetry, then derive the functions $\Phi,\Omega$ and finally comment upon 
some physical applications. We shall present
the main results as formal propositions and refer to \cite{Henk02} for details 
and the proofs. 

The defining axioms of our notion of {\em local scale invariance} from which
our results will be derived are as follows (for simplicity, in $d=1$ space
dimensions) \cite{Henk97,Henk02}. 
\begin{enumerate}
\item We seek space-time transformations with infinitesimal generators $X_n$,
such that time undergoes a M\"obius transformation
\BEQ \label{3:Moeb}
t\to t' = \frac{\alpha t + \beta}{\gamma t +\delta} \;\; ; \;\;
\alpha\delta - \beta\gamma =1
\EEQ
and we require that even after the action on the space coordinates is 
included, the commutation relations
$\left[ X_n , X_m \right] = (n-m) X_{n+m}$
remain valid. This is motivated from the fact that this condition is 
satisfied for both
conformal and Schr\"odinger invariance. 
\item The generator $X_0$ of scale transformations is
$X_0 = - t \partial_t - {\theta}^{-1} r \partial_r - {x}/{\theta}$
with a scaling dimension $x$. Similarly, the generator of time translations
is $X_{-1}=-\partial_t$. 
\item Spatial translation invariance is required.
\item Since the Schr\"odinger group acts on wave functions through a projective
representation, generalizations thereof should be expected to occur in the
general case. Such extra terms will be called {\em mass terms}. Similarly, 
extra terms coming from the scaling dimensions should be present. 
\item The generators when applied to a two-point function should yield 
a finite number of independent conditions, i.e. of the form $X_n G=0$.
\end{enumerate}

\noindent
{\bf Proposition 1}: {\it Consider the generators}
\BEQ \label{gl:X}
X_n = - t^{n+1}\partial_t 
- \sum_{k=0}^{n} \left(\vekz{n+1}{k+1}\right) 
A_{k0}r^{\theta k +1} t^{n-k} \partial_r 
- \sum_{k=0}^{n} \left(\vekz{n+1}{k+1}\right) B_{k0}r^{\theta k} t^{n-k}~~~
\EEQ
{\it where the coefficients $A_{k0}$ and $B_{k0}$ are given by the 
recurrences $A_{n+1,0} = \theta A_{n0} A_{10}$, 
$B_{n+1,0} = \frac{\theta}{n-1}\left( n B_{n0}A_{10} - A_{n0} B_{10}\right)$
for $n\geq 2$ where $A_{00}=1/\theta$, $B_{00}=x/\theta$ and in addition one of 
the following conditions holds: (a) $A_{20}= \theta A_{10}^2$ (b) 
$A_{10}=A_{20}=0$ (c) $A_{20}=B_{20}=0$ (d) $A_{10}=B_{10}=0$. 
These are the most general linear first-order operators in
$\partial_t$ and $\partial_r$ consistent with the above axioms 1. and 2. 
and which satisfy the commutation relations
$[X_{n},X_{n'}]=(n-n')X_{n+n'}$ for all $n,n'\in\mathbb{Z}$.}

\noindent
Closed but lengthy expressions of the $X_n$ for all $n\in\mathbb{Z}$ are 
known \cite{Henk02}. In order to include space translations, we set
$\theta=2/N$ and use the short-hand 
$X_n = -t^{n+1}\partial_t - a_n \partial_r - b_n$. We then define
\BEQ \label{gl:Y}
Y_m = Y_{k-N/2} = - \frac{2}{N(k+1)}\left(
\frac{\partial a_k(t,r)}{\partial r}\partial_r
+\frac{\partial b_k(t,r)}{\partial r} \right)
\EEQ
where $m=-\frac{N}{2} + k$ and $k$ is an integer. 
Clearly, $Y_{-N/2}=-\partial_r$ generates space translations. 

\noindent {\bf Proposition 2:} 
{\it The generators $X_n$ and $Y_m$ defined in eqs.~(\ref{gl:X},\ref{gl:Y}) 
satisfy the commutation relations}
\BEQ \label{XYComm_II}
\left[ X_n , X_{n'} \right] = (n-n') X_{n+n'} \;\; , \;\;
\left[ X_n , Y_m \right] = \left( n \frac{N}{2} - m \right) Y_{n+m}
\EEQ
{\it in one of the following three cases:
(i) $B_{10}$ arbitrary, $A_{10}=A_{20}=B_{20}=0$ and $N$ arbitrary. 
(ii) $B_{10}$ and $B_{20}$ arbitrary, $A_{10}=A_{20}=0$ and $N=1$.
(iii) $A_{10}$ and $B_{10}$ arbitrary, $A_{20}=A_{10}^2$,
$B_{20}=\frac{3}{2}A_{10} B_{10}$ and $N=2$.}
 
\noindent 
In each case, the generators depend on two free parameters. The physical
interpretation of the free constants $A_{10},A_{20},B_{10},B_{20}$ is still
open. In the cases (ii) and (iii), the generators close into 
a Lie algebra, see \cite{Henk02} for details. For case (i), a closed Lie
algebra exists if $B_{10}=0$. 

Turning to the mass terms, we now restrict to the projective transformations
in time, because we shall only need those in the applications later. 
It is enough to give merely the `special' generator $X_1$ which reads
for $B_{10}=0$ as follows \cite{Henk02}
\BEQ \label{gl:X1}
X_1 = -t^2\partial_t - N t r \partial_r - N x t -\alpha r^{2} \partial_t^{N-1}
- \beta r^{2} \partial_r^{2(N-1)/N} - \gamma \partial_r^{2(N-1)/N} r^{2}~~~
\EEQ
where $\alpha,\beta,\gamma$ are free parameters (the cases (ii,iii) of Prop. 2
do not give anything new). Furthermore, it turns out
that the relation $[X_1,Y_{N/2}]=0$ for $N$ integer is only satisfied in one
of the two cases (I) $\beta=\gamma=0$ which we call {\em Type~I} 
and (II) $\alpha=0$ which we call {\em Type~II}. The distinction between Type I
and Type II is very important for the physical applications. 

In both cases, all generators can be obtained
by repeated commutators of $X_{-1}=-\partial_t$, $Y_{-N/2}=-\partial_r$ and
$X_1$, using (\ref{XYComm_II}). Commutators between two generators $Y_m$ are 
non-trivial and in general only close on certain `physical' states. One might
call such a structure a {\em weak Lie algebra}.  

{\small 
Technically, these results depend on the construction of 
{\em commuting} fractional derivatives satisfying the rules 
\BEQ \label{gl:FAbl}
\partial_r^{a+b} =\partial_r^a \partial_r^b \;\; , \;\;
\left[\partial_r^a,r\right]=a\partial_r^{a-1}
\EEQ 
The standard Riemann-Liouville fractional derivative is not commutative, 
see e.g. \cite{Hilf00}. But if one uses the Hadamard definition of 
the Riemann-Liouville fractional derivatives, it is possible to generalize it 
by adding terms containing derivatives of the Dirac delta function. Then 
the relations (\ref{gl:FAbl}) can be proven \cite{Henk02}.
}

For $N=1$, the generators of both Type I and Type II reduce to those of the 
Schr\"odinger group. For $N=2$, Type I reproduces the well-known generators of 
$2D$ conformal invariance (without central charge) and Type II gives another
infinite-dimensional group whose Lie algebra is isomorphic to the one of
$2D$ conformal invariance \cite{Henk02}. The physical meaning of this
new realization remains to be understood.  

Dynamical symmetries can now be discussed as follows, by calculating the 
commutator of the generalized `Schr\"odinger-operator' $\cal S$ with $X_1$. 
We take $d=1$ and $B_{10}=0$ for simplicity (it is trival that $\cal S$
commutes with the generators $X_{-1}=-\partial_t$ and $Y_{-N/2}=-\partial_r$
of time and space translations). \\

\noindent
{\bf Proposition 3:} {\it The realization of Type I sends any solution 
$\psi(t,{r})$ with scaling dimension $x=1/2-(N-1)/N$ of the 
differential equation}
\BEQ \label{WelleI}
\mathcal{S} \psi(t,\vec{r}) =
\left( -\alpha \partial_t^N + \left(\frac{N}{2}\right)^2
{\partial}_{r}^2 \right) \psi(t,{r}) = 0
\EEQ
{\it into another solution of the same equation.}

\noindent {\bf Proposition 4:} {\it The realization
of Type II sends any solution $\psi(t,r)$ with scaling dimension
$x=(\theta-1)/2+(2-\theta)\gamma/(\beta+\gamma)$ of the differential equation}
\BEQ \label{WelleII}
\mathcal{S} \psi(t,r) =
\left( -(\beta+\gamma) \partial_t + \frac{1}{\theta^2} \partial_r^{\theta}
\right) \psi(t,r) = 0
\EEQ
{\it into another solution of the same equation.}

\noindent 
In both cases, $\cal S$ is a Casimir operator of the `Galilei'-subalgebra
generated from $X_{-1}, Y_{-N/2}$ and the generalized Galilei-transformation
$Y_{-N/2+1}$. The equations (\ref{WelleI},\ref{WelleII}) can be seen as
equations of motion of certain free field theories, where $x$ is the scaling
dimension of that free field $\psi$. These free field theories are
non-local, unless $N$ or $\theta$ are integers, respectively. Of course, 
the known dynamical symmetries of the free Schr\"odinger and Klein-Gordon
equations are recovered for $N=1$ and $N=2$. 

Phenomenological consequences will be discussed in section 4. 

\section{The case $z=2$ and the Schr\"odinger group}

The origin of local scale invariance may be understood in more
detail in the special case $z=2$. We consider the diffusion equation,
in $d$ space dimensions
\BEQ \label{7:gl:Sch-Gl}
\left(2{\cal M}\partial_t - 
\vec{\partial}_{\vec{r}}\cdot\vec{\partial}_{\vec{r}}\right)
\phi(t,\vec{r}) =0
\EEQ
For fixed $\cal M$, the Schr\"odinger group is the maximal kinematic 
invariance group \cite{Nied72} 
on the space of solutions of eq.~(\ref{7:gl:Sch-Gl}). 
It is defined by the space-time transformations ($\cal R$ is a rotation matrix
and $\alpha,\beta,\gamma,\delta \in \mathbb{R}$)
\BEQ \label{7:gl:SCH}
t \longmapsto t' = \frac{\alpha t + \beta}{\gamma t + \delta} \;\; , \;\;
\vec{r} \longmapsto \vec{r}' = \frac{{\cal R} \vec{r} + \vec{v} t + \vec{a}}
{\gamma t + \delta} \;\; ; \;\; \alpha\delta - \beta\gamma =1
\EEQ
and acts projectively on the solutions $\phi(t,\vec{r})$ \cite{Nied72}.
We now present several new results on the diffusion equation and the 
Schr\"odinger group, again in the form of propositions. For 
details and the proofs we refer to \cite{Henk03}. 

Let $\mathfrak{sch}_d$ be the Lie algebra of (\ref{7:gl:SCH}) and
$\mathfrak{conf}_{d}$ is the complexified Lie algebra of the conformal group 
in $d$ dimensions. Time-translations occur in $\mathfrak{sch}_d$ and are 
parameterized by $\beta$. We now treat the `mass' $\cal M$ not as a constant 
but as another variable, see \cite{Giul96}. The wave function 
$\phi(t,\vec{r})$ which solves eq.~(\ref{7:gl:Sch-Gl}) is replaced by 
$\psi=\psi(\zeta,t,\vec{r})$ through
\BEQ \label{gl:psi}
\phi(t,\vec{r}) = \frac{1}{\sqrt{2\pi}} \int_{\mathbb{R}}
\!\D \zeta\, e^{-\II{\cal M}\zeta} \psi(\zeta,t,\vec{r})
\EEQ
which in turn satisfies the equation of motion
\BEQ \label{gl:KG}
\left( 2\II \frac{\partial^2}{\partial \zeta \partial t} + 
\frac{\partial^2}{\partial \vec{r}^2} \right) \psi(\zeta,t,\vec{r})=0
\EEQ
While $\mathfrak{sch}_d$ acts projectively on the wave functions 
$\phi(t,\vec{r})$ (the phase factor depends on $\cal M$), the action on
$\psi(\zeta,t,\vec{r})$ is via a true representation. Furthermore, we have

\noindent {\bf Proposition 5:} {\it The Lie algebra of the maximal kinetic 
invariance group of equation (\ref{gl:KG}) is $\mathfrak{conf}_{d+2}$. 
On the space of wave functions $\psi$ 
defined through eq.~(\ref{gl:psi}) one has the embedding}
\BEQ
\mathfrak{sch}_d\subset\mathfrak{conf}_{d+2}
\EEQ

\begin{figure}[th]
\centerline{\epsfxsize=1.5in\ \epsfbox{
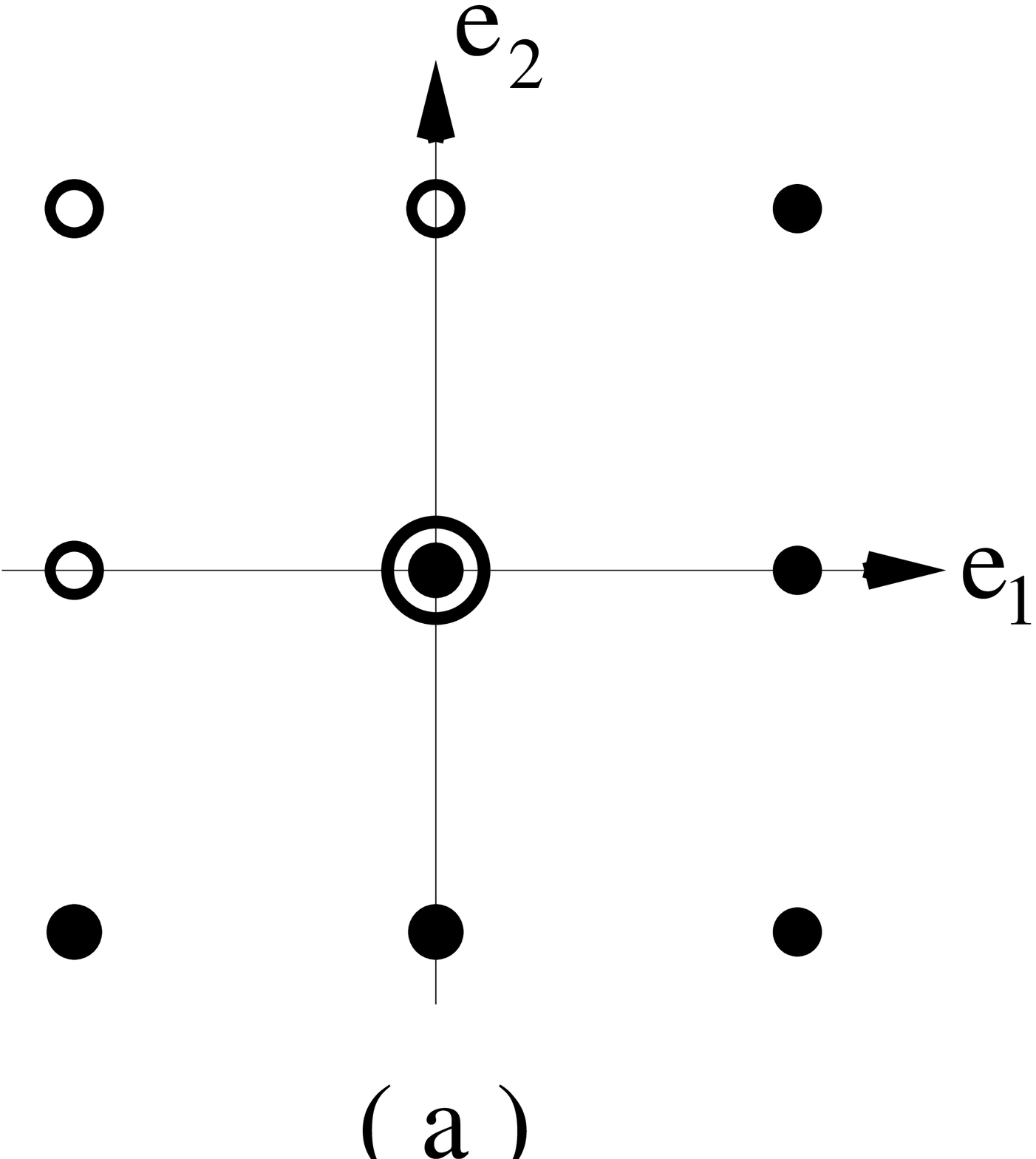} ~
\epsfxsize=1.5in\epsfbox{
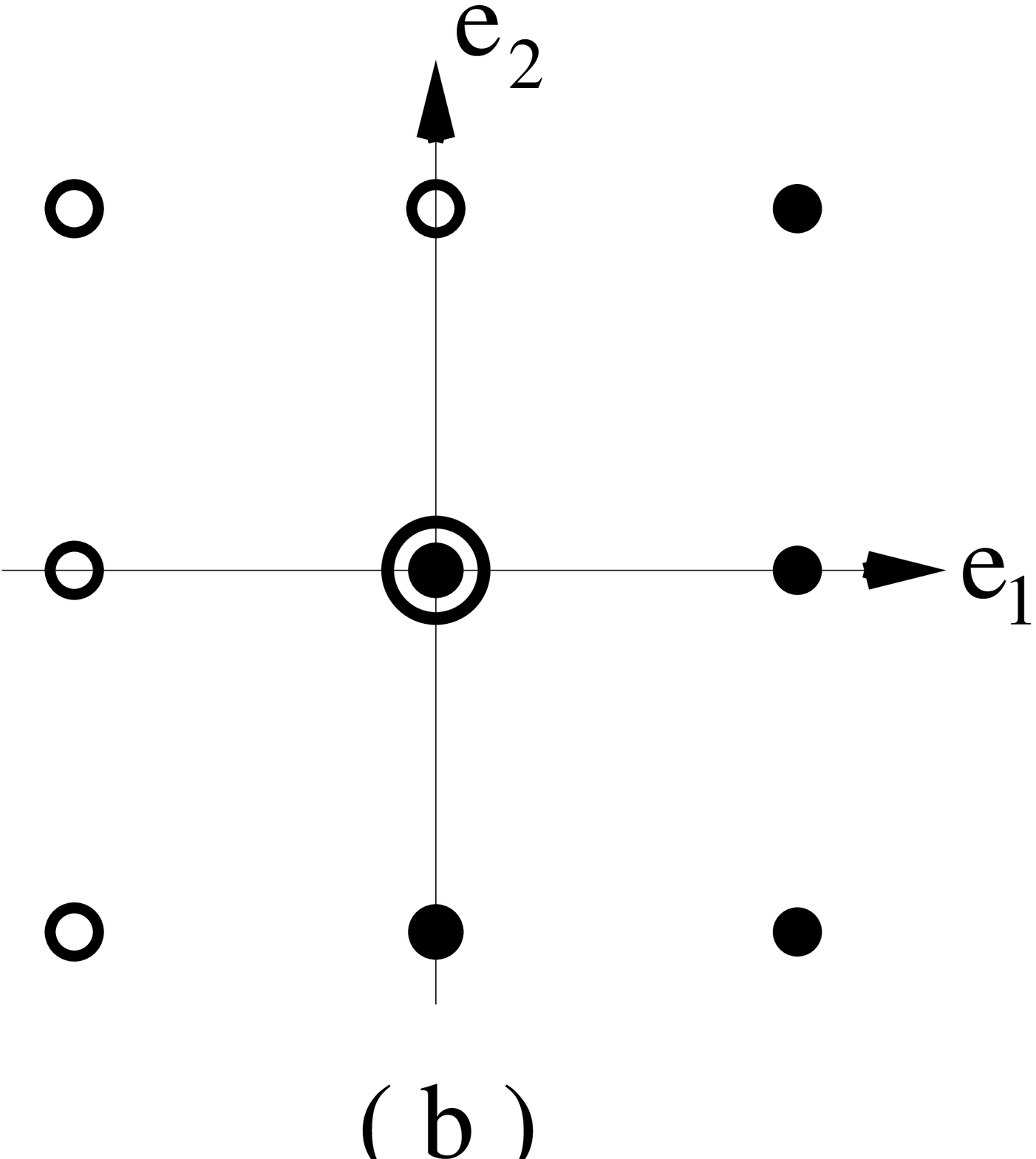} ~
\epsfxsize=1.5in\epsfbox{
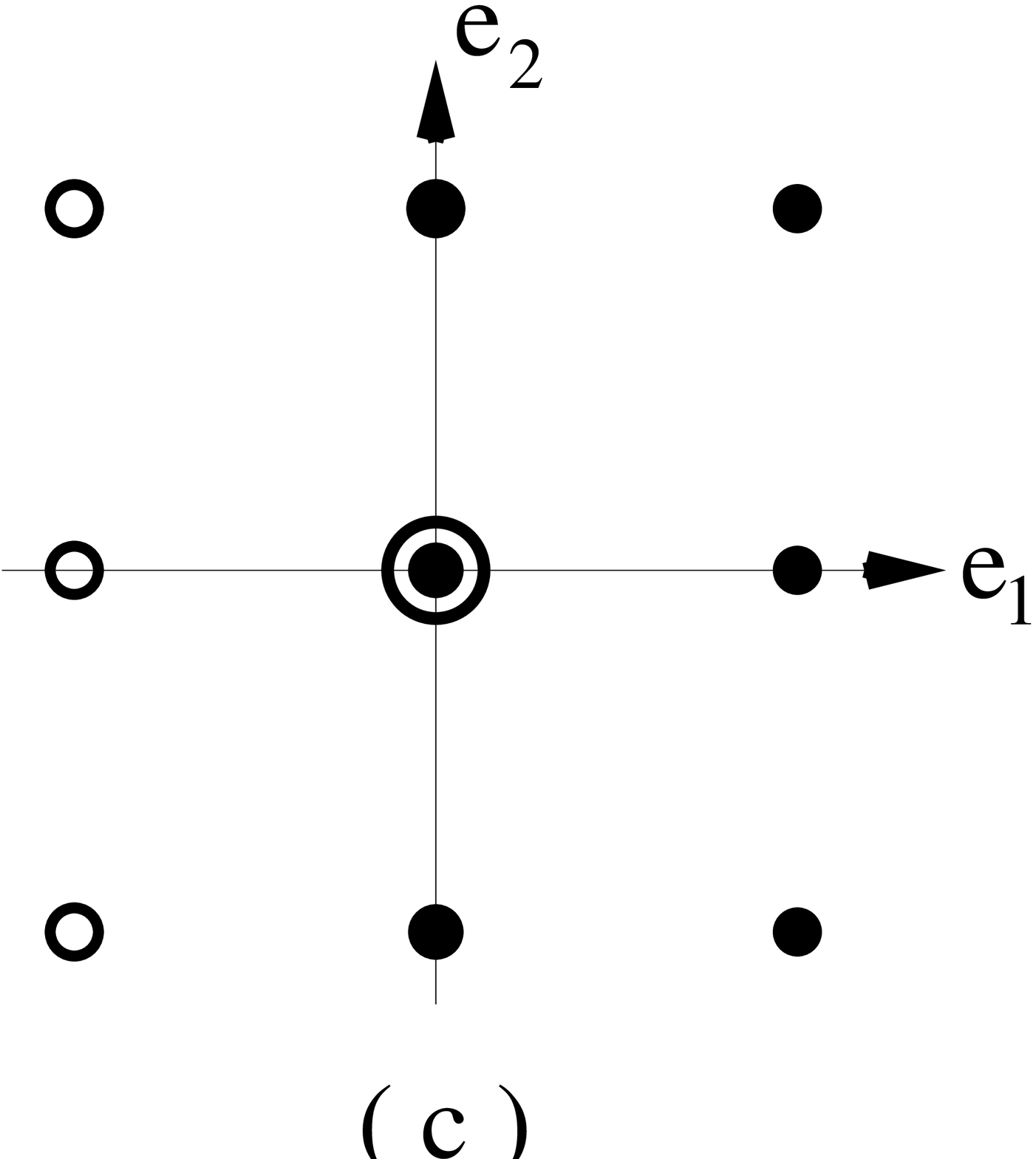}
}
\caption[Root space]{Root space of the complexified conformal Lie algebra
$\mathfrak{conf}_3$, indicated by the full and the open points. The double
circle in the center denotes the Cartan subalgebra. The generators belonging 
to the three non-isomorphic parabolic subalgebras \protect{\cite{Henk03}} 
are indicated by the full points, namely
(a) $\wit{\mathfrak{sch}}_1$, (b) $\wit{\mathfrak{age}}_1$ and
(c) $\wit{\mathfrak{alt}}_1$. 
\label{Bild6}}
\end{figure}

\noindent It is now of interest to study the subalgebras of
$\mathfrak{conf}_{d+2}$. In particular,  
the parabolic subalgebras of $\mathfrak{conf}_{d+2}$ can be classified
\cite{Henk03} and we obtain several
new subalgebras, called $\wit{\mathfrak{age}}$ or 
$\wit{\mathfrak{alt}}$. For the case of one spatial dimension $d=1$, 
we illustrate in figure~\ref{Bild6} 
their definition through the root space of $\mathfrak{conf}_3\cong B_2$. 
To each point of the diagram corresponds a generator of 
$\mathfrak{conf}_3$. For example, the generator of dilatations $X_0$ is
one of the two generators sitting at the origin (the Cartan subalgebra 
$\mathfrak{h}$), the generator $X_{-1}$ of time translations has the
coordinates $(-1,-1)$ and sits in the lower left corner of 
figure~\ref{Bild6}abc and the 
generator $Y_{-1/2}$ of space translations has the coordinates
$(0,-1)$. Subalgebras of $\mathfrak{conf}_3$ are identified by
considering all convex subset in this diagram. 

Two of these parabolic subalgebras (figure~\ref{Bild6}bc) 
still contain the generator
for the dilatations $t\to b^2 t, \vec{r}\to b \vec{r}$ (which is in the
Cartan subalgebra of $\mathfrak{conf}_3$) but do {\em not} contain 
time-translations anymore. In particular, 
$\wit{\mathfrak{age}}_1$ and $\wit{\mathfrak{alt}}_1$ 
are candidates for a dynamic symmetry algebra of ageing systems. 

While we have worked here in units such that the `speed of light' $c=1$, it
is possible to reintroduce $c$ explicitly into the equations of motion such
as eq.~(\ref{7:gl:Sch-Gl}). It had been claimed in the past that 
$\mathfrak{sch}_d$ could be obtained from a group contraction from 
$\mathfrak{conf}_{d+1}$ in the
`non-relativistic limit' $c\to\infty$ \cite{Baru73}. This limit procedure
can be made precise in terms of the wave function (\ref{gl:psi}) which 
satisfies a $(d+2)$-dimensional massless Klein-Gordon equation (\ref{gl:KG}). 
We obtain the result 

\noindent {\bf Proposition 6:} {\it The non-relativistic limit $c\to\infty$
of the conformal group acting on the wave function $\psi$ defined through
(\ref{gl:psi}) leads to the map}
\BEQ
\mathfrak{conf}_3 \to \wit{\mathfrak{alt}}_1
\EEQ

\noindent Although $\wit{\mathfrak{alt}}_1$ and $\wit{\mathfrak{sch}}_1$ have 
the same dimension, they are not isomorphic, see figure~\ref{Bild6}ac. 

The dynamical symmetries of the free diffusion equation
may be related to the symmetries of the action $S$ of the associated free-field
theory. This action may be constructed in the standard fashion
following the lines of Martin-Siggia-Rose theory, see e.g. \cite{Card96} and 
references therein. Working with the wave function $\phi=\phi(t,\vec{r})$, we
introduce a compound vector $\vec{\eps}=(\delta t,\delta \vec{r})$ for small
coordinate changes. Let $\eta$ be the phase which arises in the transformation
of the wave function $\phi$ under local coordinate transformations. For local
theories we expect the following transformation of the action
\BEQ \label{gl:local}
\delta S= \int\!\D t \, \D\vec{r}\: 
\left( T_{\mu\nu}\partial^{\mu}\vec{\eps}^{\nu}
+ J_{\mu}\partial^{\mu}\eta\right) 
+ \int_{t=0}\!\D\vec{r}\: \left( U_{\nu}\vec{\eps}^{\nu} + V \eta\right)
\EEQ
where $T_{\mu\nu}$ is the conserved energy-momentum tensor and $J_{\mu}$
is the conserved probability current. The second term 
contains the contributions
from the initial conditions. This local form permits the derivation of
Ward identities and we obtain

\noindent
{\bf Proposition 7:} {\it Consider an ageing system described 
by a local action $S$ transforming according to (\ref{gl:local}) 
and which is invariant under spatial translations, 
phase shifts, Galilei transformations and dilatations with $z=2$. 
Then $\delta S=0$ under the action of the special Schr\"odinger 
transformation $X_1$.\footnote{$X_1$ is the generator of the
`special' transformation parameterized by $\gamma$ in eq.~(\ref{7:gl:SCH}).}}

\noindent This illustrates the conceptual importance of the requirement of
Galilei invariance, besides dynamical scaling with $z=2$, for full local
scale invariance to hold. We stress that time-translation invariance is not
required.

Finally, we mention that there exist infinite-dimensional Lie algebras 
which contain $\mathfrak{sch}_d$
as subalgebras. For example, the Schr\"odinger group (\ref{7:gl:SCH}) is a
subgroup of the group defined by the transformations $t\to t'$ and 
$\vec{r}\to\vec{r}'$ where \cite{Henk03}
\BEQ
t' = \beta(t) \;\; , \;\; \vec{r}' = \vec{r} \sqrt{\dot{\beta}(t)\:}
\;\; \mbox{\rm ~~ or else ~~}
t' = t \;\; , \;\; \vec{r}' = \vec{r} - \vec{\alpha}(t) 
\EEQ
and $\beta$ and $\vec{\alpha}$ are arbitrary functions. 
Whether this has a bearing on the non-equilibrium behaviour of
spin systems is still open.

\section{Scaling functions}

{}From a physical point of view, the wave equations 
(\ref{WelleI},\ref{WelleII}) discussed in section 2 suggest that the 
applications of Types I and II are very different. Indeed, eq.~(\ref{WelleI})
is typical for equilibrium systems with a scaling anisotropy introduced through
competing uniaxial interactions. Paradigmatic cases of this are so-called
Lifshitz points which occur for example in magnetic systems when an
ordered ferromagnetic, a disordered paramagnetic and an incommensurate phase
meet (see \cite{Dieh02} for a recent review). On the other hand, 
eq.~(\ref{WelleII}) is reminiscent of a Langevin equation which may describe
the temporal evolution of a physical system, with a dynamical exponent
$z=\theta$. In any case, causality requirements
can only be met by an evolution equation of first order in $\partial_t$. 

We now give the scaling functions $\Phi,\Omega$ in eq.~(\ref{gl:skala}) from
the assumption that the two-point function $G$ transforms covariantly under 
local scale transformations.

\noindent
{\bf Proposition 8:} \cite{Henk02} 
{\it Local scale invariance implies that for Type I with $\theta=2/N$, 
the function $\Omega(v)$ must satisfy}
\BEQ \label{gl:Omega}
\left( \alpha \partial_v^{N-1} - v^2 \partial_v - N x \right) \Omega(v) = 0
\EEQ
{\it together with the boundary conditions $\Omega(0)=\Omega_0$ and 
$\Omega(v)\sim\Omega_{\infty} v^{-Nx}$ for $v\to\infty$. For Type II, we have}
\BEQ \label{gl:Phi}
\left( \partial_u +\theta(\beta+\gamma)u\partial_u^{2-z} 
+2\theta(2-z)\gamma\partial_u^{1-z}\right)\Phi(u)=0
\EEQ
{\it with the boundary conditions $\Phi(0)=\Phi_0$ and 
$\Phi(u)\sim\Phi_{\infty}u^{-2x}$ for $u\to\infty$.} 

\noindent 
Here $\Omega_{0,\infty}$ and $\Phi_{0,\infty}$ are constants. The ratio
$\beta/\gamma$ turns out to be universal and related to $x$. 
{}From the linear differential equations (\ref{gl:Omega},\ref{gl:Phi}) 
the scaling functions $\Omega(v)$ and $\Phi(u)$ can be found
explicitly using standard methods \cite{Henk02}. 

\section{Application to Lifshitz points}

\begin{figure}[ht]
\centerline{\epsfxsize=3.65in\epsfbox
{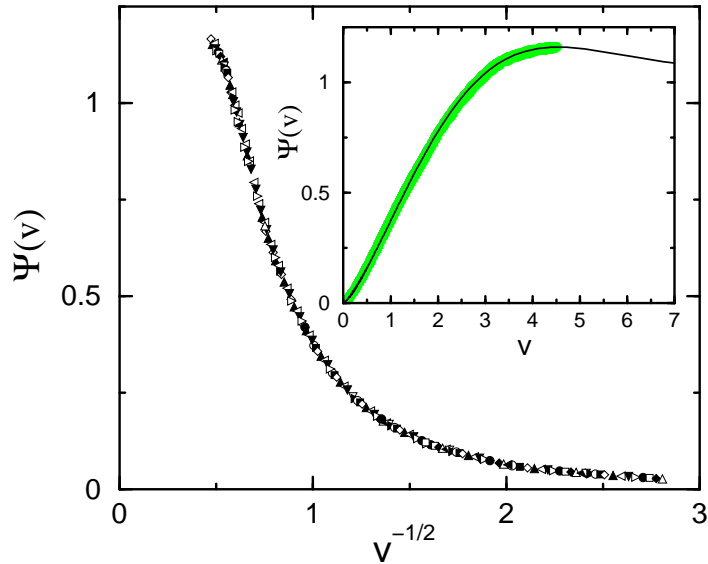}
}
\caption{Scaling function $\Psi(v)$ of the spin-spin correlator at the
Lifshitz point in the $3D$ ANNNI model. 
In the main plot the scaling of selected 
data for several values of $r_{\perp}$ is shown. In the inset, the full set of
numerical data (gray points) is compared to the solution of
eq.~(\ref{gl28}). The data are from {\protect \cite{Plei01}}.  
\label{Bild1}}
\end{figure}

Given these explicit results, the idea of local scale invariance can be
tested in specific models. Indeed, the predictions for $\Omega(v)$ coming from
Type I with $N=4$ nicely agree with cluster Monte Carlo data 
for the spin-spin and energy-energy correlators of the $3D$ ANNNI model at its
Lifshitz point. The ANNNI model is defined in terms of 
Ising spins $\sigma=\pm 1$ on
a hypercubic lattice. The Hamiltonian is
\BEQ
{\cal H} = -\sum_{xyz} \sigma_{xyz}\left[ 
\sigma_{xy(z+1)}+\sigma_{x(y+1)z}+\sigma_{(x+1)yz}\right]
+\kappa \sum_{xyz} \sigma_{xyz}\sigma_{xy(z+2)}
\EEQ
where $\kappa$ is a constant. 
Extensive simulations using a generalization \cite{Plei01,Henk01b} 
of the Wolff algorithm \cite{Wolf89} and a 
new method \cite{Ever01} to reduce finite-size effects in the estimates 
of correlation functions led to a considerable
improvement in the precision of the estimation of the location of the 
Lifshitz point, viz. $\kappa_L=0.270(4)$ and $T_L=3.7475(50)$ \cite{Plei01}.
The Lifshitz-point exponents 
$\alpha_L=0.18(2), \beta_L=0.238(5),\gamma_L=1.36(3)$, respectively, 
of the specific heat, the magnetization and the susceptibility were 
obtained \cite{Plei01}.\footnote{The fact that the scaling relation
$\alpha_L+2\beta_L+\gamma_L=2$ is satisfied up to $\approx 0.8\%$ gives
an {\it a posteriori} estimate of the quality of the numerical data.}
These values are also in good agreement with the results of a careful two-loop
study of the $3D$ ANNNI model within renormalized field theory which
gives $\alpha_L=0.160$, $\beta_L=0.220$ and $\gamma_L=1.399$ \cite{Shpo01}. 
In figure~\ref{Bild1}, we show the scaling function 
$\Psi(v) := v^{\zeta} \Omega(v)$ of the spin-spin correlator. Here 
$\zeta=(2/\theta)(d_{\perp}+\theta)/(2+\gamma_L/\beta_L)$, where $d_{\perp}=2$ 
is the number of transverse dimensions without
competing terms and $\theta\simeq 1/2$ within the numerical accuracy of the
data (see \cite{Plei01,Shpo01,Henk02} for a fuller discussion of this point). 
We see the expected collapse of the data which establishes scaling. In the
inset, we compare the data with the prediction of local scale invariance. 
If $\theta=2/N=1/2$, we have $N=4$ and expect from proposition 8 that
$\Omega(v)$ satisfies
\BEQ \label{gl28}
\alpha \frac{\D^3 \Omega(v)}{\D v^3} - v^2 \frac{\D \Omega(v)}{\D v}
- 4x v \Omega(v) = 0
\EEQ
The are two independent solutions which satisfy the required boundary
conditions, such that the form of $\Omega(v)$ depends on a single
universal parameter $p$. Its value can in turn be estimated by
considering universal ratios of moments of $\Omega(v)$, see \cite{Plei01}
for details. In the inset of figure~\ref{Bild1}, we find perfect
agreement with the prediction of local scale invariance. 

Replacing the Ising model spins by so-called spherical spins 
$s_{\vec{r}}\in\mathbb{R}$ such that 
$\sum_{\vec{r}\in\Lambda} s_{\vec{r}}^2 = {\cal N}$,
where $\cal N$ is the number of sites of the lattice, one obtains the 
exactly solvable ANNNS model. At its Lifshitz point, 
the spin-spin correlator agrees
exactly with the prediction of local scale 
invariance \cite{Frac93,Henk97,Henk02}.\footnote{The ANNNS model may be
generalized such that Lifshitz points of second order appear as the endpoints
of lines of Lifshitz points. This allows for a test of local scale invariance
for $N=6$ \cite{Henk02}.}

On the other hand, the predictions of 
Type II have been tested extensively in the context of ageing ferromagnetic 
spin systems which are going to be described in the next section.

\section{Application to ageing}

Time translation invariance does {\em not} hold in ageing systems. 
The simplest way to do this is to remark that the 
Type~II-subalgebra spanned by $X_0, X_1$ and the $Y_m$ leaves the initial 
line $t=0$ invariant, see (\ref{gl:X1}). Therefore the autoresponse function
$R(t,s)$ is fixed by the two covariance conditions $X_0 R = X_1 R=0$. 
Solving these differential equations leads to 

\noindent
{\bf Proposition 9:} \cite{Henk01,Henk02}
{\it For a statistical non-equilibrium system which 
satisfies local scale invariance the autoresponse function $R(t,s)$ takes the
form}
\BEQ \label{gl:R}
R(t,s) = r_0 \left( t/s \right)^{1+a-\lambda_R/z} \left( t-s\right)^{-1-a}
\;\; , \;\; t > s~~~~
\EEQ  
{\it where $a$ and $\lambda_R/z$ are the non-equilibrium exponents defined 
in section 1 and $r_0$ is a normalization constant.} 

\noindent The functional form of $R$ 
is completely fixed once the exponents
$a$ and $\lambda_R/z$ are known. Similarly, 
$R(t,s;\vec{r})=R(t,s) \Phi\left(r (t-s)^{-1/z}\right)$ gives the 
spatio-temporal response, with the
scaling function $\Phi(u)$ determined by (\ref{gl:Phi}) \cite{Henk02}. 
In the $z=2$ special case the above result takes a particularly simple form.

\noindent 
{\bf Proposition 10:} \cite{Henk94,Henk03}
{\it For an ageing system with dynamical exponent $z=2$ and which is invariant
under $\mathfrak{age}_d$, the two-time spatio-temporal response function is}
\BEQ \label{gl:RRz2}
R(t,s;\vec{r}) = R(t,s) \exp\left[ -\frac{\cal M}{2} \frac{\vec{r}^2}{(t-s)}
\right]
\EEQ
{\it where $R(t,s)$ is given by eq.~(\ref{gl:R}).}

\noindent The form of the two-point function does not change if we use
$\mathfrak{alt}_d$ instead. 

Ageing systems are often described in terms of Langevin equations
of the form
\BEQ
\dot{\phi}(t) = - \frac{\delta {\cal H}}{\delta \phi(t)} + \eta(t)
\EEQ
where $\eta(t)$ is assumed to be a Gaussian noise with mean zero and variance
$\langle \eta(t)\eta(t')\rangle = 2T \delta(t-t')$. In the context of
Martin-Siggia-Rose theory, it can be shown that response functions can be
written as a correlator $R(t,s)=\langle \phi(t)\wit{\phi}(s)\rangle$ of the
order parameter $\phi$ and its associated response field $\wit{\phi}$. On the
other hand, to each solution $\phi(t,\vec{r})$ with positive mass ${\cal M}>0$
there exists a conjugate solution $\phi^*(t,\vec{r})$ which satisfies
the diffusion equation with $\cal M$ replaced by $-{\cal M}$ and which can be
said to have negative mass. 

On the other hand, if we work instead with the wave function 
$\psi(\zeta,t,\vec{r})$ as defined in eq.~(\ref{gl:psi}), we can use the
embedding $\mathfrak{sch}_d\subset \mathfrak{conf}_{d+2}$ and the form of the
two- and three-point functions is fixed by conformal invariance. For ageing
systems, time-translation invariance does not hold and we must restrict to
those parabolic subalgebras which do not contain $X_{-1}$. We are interested
in the linear response functions
\BEA
R_2(t_1,s) &=& \frac{\delta \langle \phi(t_1)\rangle}{\delta h(s)} 
= \langle \phi(t_1)\wit{\phi}(s)\rangle 
\\
R_3(t_1,t_2,s) &=& \frac{\delta \langle \phi(t_1)\phi(t_2)\rangle}{\delta h(s)} 
= \langle \phi(t_1)\phi(t_2)\wit{\phi}(s)\rangle 
\EEA
Instead, we may also calculate them using the functions $\psi(\zeta,t,\vec{r})$
and then find

\noindent
{\bf Proposition 11:} \cite{Henk03} {\it For an ageing system with $z=2$
and which is invariant under $\mathfrak{age}_d$, the two- and three-time
spatio-temporal response functions may be obtained from}
\BEA
R_2(t_1,s) &=&  \langle \phi(t_1)\phi^*(s)\rangle
\\
R_3(t_1,t_2,s) &=&  
\langle \phi(t_1)\phi(t_2)\phi^*(s)\rangle\;\;
\EEA
{\it such that the causality conditions $t_1>s$ and $t_2>s$ are automatically
satisfied.}

\noindent This suggests the identification of the response field 
\BEQ
\wit{\phi}=\phi^*
\EEQ 
with the conjugate solution of the Schr\"odinger equation.
Explicit expressions for these response functions are known.  

Finally, Galilei invariance holds {\it de rigueur} only 
for a vanishing temperature $T=0$ and
in principle all predictions made on ageing systems so far are restricted to
that case.\footnote{At $T=0$, the ageing behaviour is determined from the
properties of the initial state.} However, constructing the free-field 
Martin-Siggia-Rose action it can be
seen through the Wick theorem that the perturbative series in $T$ terminates
after the first term. 

\noindent 
{\bf Proposition 12:} \cite{Pico04} {\it For an ageing Gaussian 
system with $z=2$
and which is invariant under $\mathfrak{age}_d$ and at a temperature 
$T<T_c$, the response function $R(t,s)$ agrees to all orders in $T$ with
eq.~(\ref{gl:R}).}

\noindent This argument is expected to work in the low-temperature phase,
where $T$ flows to zero under the action of the renormalization group
\cite{Bray94} so that perturbation theory should be applicable. Direct 
evidence in favour of this is given below. The independence of the
form of $R(t,s)$ on the (short-ranged) initial conditions is all the more
remarkable since for instance the correlators do depend on them \cite{Pico04}. 
On the other hand, {\em at} $T=T_c$ thermal fluctuations and fluctuations of the
initial state become simultaneously important. Remarkably enough, there 
exists good numerical evidence that eq.~(\ref{gl:R}) even holds at $T=T_c$
in the $2D$ and $3D$ Glauber-Ising model \cite{Henk01}. 
However, a two-loop renormalization
group calculation of the O($n$) field theory gives a numerically small 
correction to eq.~(\ref{gl:R}) \cite{Cala03}. The understanding of the
r\^ole of temperature on the validity of Galilei invariance in field theory 
will require more work.

A quantitative test of the prediction (\ref{gl:R}) is best carried out through
the consideration of an integrated response function, such as the
thermoremanent magnetization 
\BEQ
M_{\rm TRM}(t,s)=h\int_{0}^{s} \!\D u\,R(t,u) = r_0\, s^{-a} f_M(t/s)
\EEQ
where the system is quenched in a small (random \cite{Barr98}) 
magnetic field $h$ which is turned off
after the waiting time $s$ has elapsed. The magnetization is then measured at
a later time $t$. Local scale invariance furnishes, via (\ref{gl:R}), 
an explicit predictions for the scaling function $f_M(x)$. In certain
cases there may be sizeable correction terms to this scaling form, see
\cite{Henk02a,Corb02,Henk03b,Henk02b,Zipp00} for details.

\begin{figure}[t]
\centerline{\epsfxsize=3.5in\ \epsfbox{
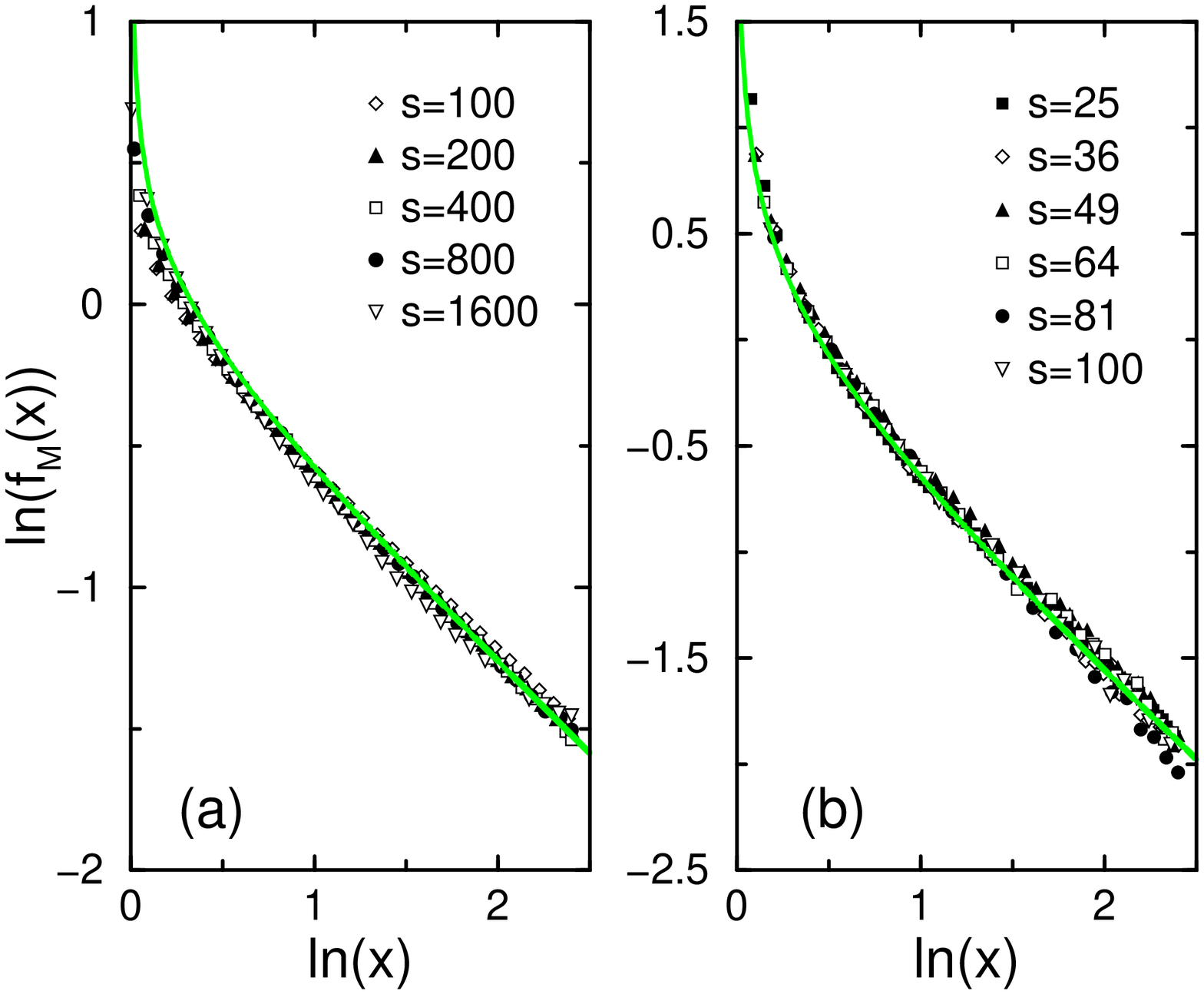} ~
\epsfxsize=1.88in\epsfbox{
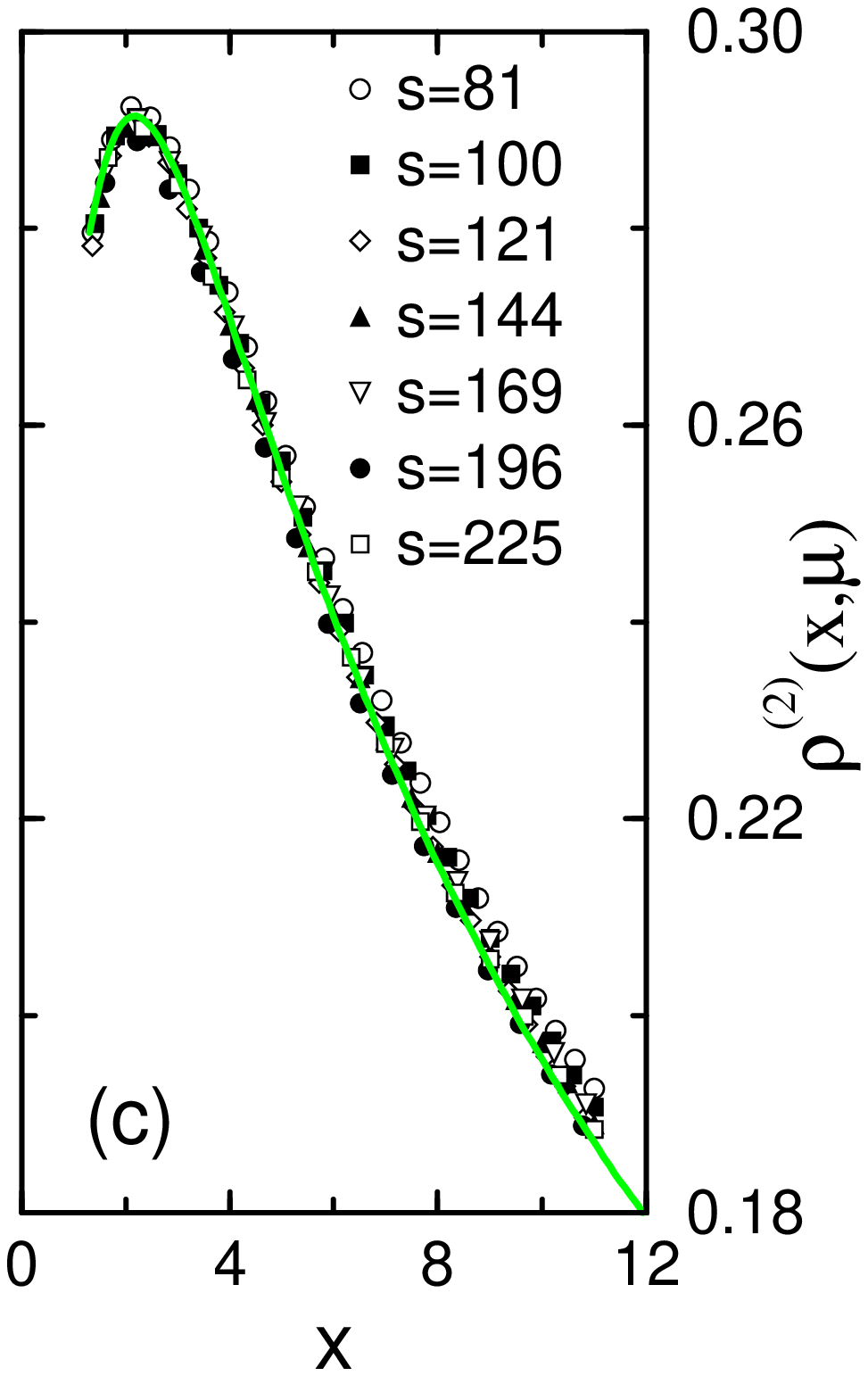}
}
\caption[Space-time response]{Scaling form of the integrated magnetic response
in the Glauber-Ising model as a function of $x=t/s$ below criticality. The
symbols correspond to different waiting times $s$. The integrated autoresponse 
is shown (a) in $2D$ at $T=1.5$ and (b) in $3D$ at $T=3$. An example of the 
integrated spatio-temporal response in $2D$ at $T=1.5$ and with $\mu=2$ is 
shown in (c). The full curves are obtained from (\ref{gl:R}). 
After \protect{\cite{Henk02b}}.
\label{Bild5}}
\end{figure}

We now consider ageing in the Glauber-Ising model (\ref{7:gl:heat}), quenched
to a temperature $T<T_c$ from a fully disordered initial state. Then the
exponents $a=1/z=1/2$ and $\lambda_R\simeq 1.26$ and $1.6$
in $2D$ and $3D$, respectively, are known. 
In figure~\ref{Bild5}ab, the scaling function
$f_M(x)$, as obtained from large-scale simulations, 
is shown for several values of
the waiting time $s$. In both two and three dimensions, a nice scaling behaviour
is found and the form of the scaling function agrees very well with the
prediction from eq.~(\ref{gl:R}). 

For $z=2$, we have seen in section 3 the importance of Galilei invariance as
the second building block, besides dynamical scaling, of the local scale
invariance of systems whose action $S$ transforms according to
eq.~(\ref{gl:local}). It is therefore important to test Galilei 
invariance directly, which can be done by considering the
spatio-temporal response $R(t,s;\vec{r})$ and testing it for the form 
eq.~(\ref{gl:RRz2}) \cite{Henk02b}.
We first fix the mass $\cal M$ by studying how the
integrated response $\int_{0}^{s}\!\D u\, R(t,u;\vec{r})$ depends on
$r^2/s$ for a single fixed value of $x=t/s$. Next, a demanding test of 
the $\vec{r}$-dependence of $R(t,s;\vec{r})$ can be performed 
by measuring the spatio-temporally integrated response
\BEQ
\int_{0}^{s} \!\D u\int_{0}^{\sqrt{\mu s}} \!\D r\, r^{d-1} R(t,u;\vec{r}) 
\sim s^{d/2-a} \rho^{(2)}(t/s,\mu)
\EEQ 
where $\mu$ is a control parameter. We stress that the scaling function 
$\rho^{(2)}$ 
does not contain any free non-universal parameter at all \cite{Henk02b}. As an
example, we compare in figure~\ref{Bild5}c data from $2D$ taken with $\mu=2$
with eq.~(\ref{gl:R}). Besides the expected scaling, the functional form of
the scaling function neatly follows the prediction. We stress that the
position, the height and the width of the maximum of $\rho^{(2)}$ in 
figure~\ref{Bild5}c are completely fixed. Similar results have been
obtained for other values of $\mu$ and in $3D$ as well. This provides strong
evidence that eq.~(\ref{gl:RRz2}) is exact, at least in this model 
\cite{Henk02b}. This is the first time that Galilei invariance in an ageing
system has been directly confirmed. It is remarkable that Galilei invariance,
which after all was used with a generator which takes the form found for
a free particle, is confirmed for a theory which is certainly {\em not} a
free-field theory. 

The prediction (\ref{gl:R}) for the autoresponse has been confirmed  
in several physically distinct exactly solvable systems undergoing ageing, 
including several variants of the kinetic spherical model, the voter
model and Brownian motion, 
see \cite{Godr02,Pico02,Cugl94b,Henk01,Cann01,Cala02,Dorn02} 
and references therein. 
Where applicable, these models also confirm the form (\ref{gl:RRz2}) of the
spatio-temporal response. 
 
Taken together, these confirmations suggest that (\ref{gl:R}) should hold
independently of 
\begin{enumerate}
\item the value of the dynamical exponent $z$ 
\item the spatial dimensionality $d>1$ 
\item the numbers of components of the order parameter and
the global symmetry group 
\item the spatial range of the interactions 
\item the presence of spatially long-range initial correlations 
\item the value of the temperature $T$ 
\item the presence of weak disorder
\end{enumerate} 
This provides strong evidence that local scale invariance is indeed realized
as a dynamical symmetry in ageing systems. 

\section{Conclusion}

Motivated by the observation that dynamical or strongly anisotropic 
scaling occurs in many physically relevant situations, we have made the 
hypothesis of dynamical scaling the starting point of our study. Since
conformal transformations may be seen as a sort of space-dependent
local scale transformations, we have asked ourselves whether an analogous
extension might be possible for any given value of the dynamical exponent
$z$. Indeed, we have shown that infinitesimal transformations of this kind
can be explicitly constructed. It turns out that there exist two distinct
realizations of these local scale transformations (Type I and Type II) 
which lead to very different physical consequences. 

These purely formal considerations appear to have a bearing on 
physics, since the predictions of local scale invariance for some two-point
functions could be explicitly verified in concrete physical models,
which describe either Lifshitz points (Type I) or ageing systems (Type II).
Several of these models do not reduce to a free-field theory. We have also
indicated in the text some of the open problems which remain and which we hope 
might be resolved at a later time. 

We hope that in the future, the principle of local scale invariance might
become useful in order to derive hitherto unexplained properties of 
dynamic (or strongly anisotropic) phase transitions. For example, is there
a way to predict the values of the exponents which we have used here as
externally given parameters~?\\

This work was supported by CINES Montpellier (projet pmn2095) and the 
Bayerisch-Franz\"osisches Hochschulzentrum (BFHZ). 
MH thanks the Centro de F\'{\i}sica Te\'orica e Computacional (CTFC) of the
Universidade de Lisboa for warm hospitality, where this work was 
finished.

\end{document}